\documentstyle[prb,aps]{revtex} 
\begin{document} 
\draft 
\title {Signatures of
 Wannier-Stark and surface states\\ in electron tunneling and related 
phenomena:\\ Electron transmission through a
tilted band } 
\author{Alexander \ Onipko and  Lyuba \ Malysheva$^\dagger$}
\address{
Department of Physics, Uppsala University, Box 530, S-751 21 
Uppsala, Sweden\\
\footnote{Also the address for communication}Department of Physics, IFM, 
Link\"oping 
University, S-581 83 Link\"oping, Sweden
}
\address{$^\dagger$Bogolyubov Institute for Theoretical Physics, 
Kiev, 03143, Ukraine
}
\date{\today} \maketitle
\begin{abstract}

Predicted by Wannier in 1960 band states quantization in a constant electric
field, $E_n$ = const $\pm n {\cal E}$, where $n =0$, 1, 2, ..., and ${\cal E}$
is proportional to the strength of electric field [this kind of spectrum is
commonly referred as the Wannier-Stark ladder (WSL)], implies that the
probability of tunneling through a tilted band should have ${\cal E}$ spaced
peaks, at least, under the weak coupling of the band states to the source and
drain electrodes.  It has been shown, however, (Phys. Rev. B {\bf 63},
..., 2001), that the appearance of the canonical WSL is preceded by WSLs with
other level spacing, namely, ${\cal E}_{m'/m}/(1-2m'/m)$, where  $m$ and 
$m'< m/2$ are positive integers specifying certain applied voltage. Here we
show that canonical and noncanonical WSLs, in addition to different peak
spacing in the transmission spectrum, have other pronouncedly distinctive
features.  As an example, for the former, the peak and valley tunneling
probability decays exponentially with the increase of applied voltage.  The
corresponding exponents are given by the sum and difference of two
Fowler-Nordheim-type exponents implying an anomalous increase of the
peak-to-valley ratio. These and other signatures of extended states (es), 
surface 
localized states (sls), and Wannier-Stark (WS) states in the through tilted band 
transmission spectrum are discussed on the basis of (derived for the first time) 
nearly exact explicit expressions of es-, sls-, and WS states assisted tunneling 
probability.

\end{abstract}

\newpage
\section{Introduction}

Introduced by Wannier in 1960,\cite{1,2} the concept of Wannier-Stark ladder 
(WSL), associated with the spectrum of electron band states in a constant 
electric field, gave rise to an unobservable literature of discussions aimed at 
the development of in-depth understanding of electric field effects in solids. 
In nineties, this activity has resulted in 
several reports on experimental observation of the Wannier-Stark (WS) effect, 
Bloch 
oscillations, and related coherent phenomena,\cite{3,4,5} appearance of new 
ideas (such as, e.g., dynamic localization\cite{6} and quasi-energy band 
collapse 
in 
time-dependent electric fields,\cite{7,8,9}) and new possibilities 
to study static and time dependent field effects in optically driven 
lattices.\cite{10,11,12,13,14,15}

In its canonical form, the WSL represents a spectrum of evenly spaced electronic 
levels $E_n$ =
const $\pm n {\cal E}$, where $n =0$, 1, 2, ..., and ${\cal E}$ (the level
spacing proportional to the strength of electric field) is equal to the Plank
constant times Bloch oscillation frequency.  In the tight-binding description,
this kind of spectrum is readily expected,\cite{16} if the constant shift ${\cal
E}$ of electron on-site (atomic) energy along the electric field is comparable
with the width of the parent zero-field band of Bloch electron states.  Such an
extreme case of WSL, though is easy to think about, is of little, if any,
physical interest, since in most cases, the band state spectrum emerges as a
result of delicate interplay of electric field and interatomic interaction
effects, where the methodology of perturbation theory is inappropriate.

In early seventies, it was realized that the tight-binding model presents a
unique opportunity to treat field effects in solids on rigorous basis, since its
Hamiltonian admits obtaining the exact formal solution of the Schr\"odinger
equation\cite{17,18,19,20} and hence, the Green function problem,\cite{21,22,23}
without any restriction on the field parameter.  A number of useful results has
been obtained in this way,\cite{17,18,19,22,23,24} in particular, that in finite
systems, some minimal voltage is required to open the WS band.\cite{19} Unlike
the Wannier model, which refers to the bulk electron states only, the
field-induced surface states also have been brought to light.  It was shown that
eigenenergies near the spectrum edges behave similarly to zeros of the Airy
function.\cite{18,19,24} Recently,\cite{25} within the same model, new
regularities for the bulk and surface electron states influenced by a constant
electric field were predicted.  The bulk levels, for instance, have been shown
to form (under certain voltages) WSLs with noncanonical level spacing $q{\cal
E}$, where $q$ ($>1$) can be an integer, as well as a fractional number.  Thus
the concept of WSL appeared even more rich and fertile physically, than was
commonly thought.

Next step to be undertaken is to specify such physical quantities, which
allow one to observe distinctions between the true Wannier levels and the 
others, for instance, the levels that correspond to {\it extended states} (es) 
and form noncanonical WSLs, and those corresponding to {\it
surface localized states} (sls), whose spacing is governed by three different 
laws.\cite{25}  Here, in focus is the probability of tunneling
(or transmission) of electrons through a tilted tight-binding band. This 
quantity is of relevance to the phenomenon of electrical breakdown of ultra-thin 
dielectric layers and a number of other electric field effects in solids. 
The energy dependence of electron transmission can be directly probed by means 
of 
the 
ballistic-electron-emission microscopy technique.\cite{26} And properties of 
through tilted band tunneling are intimately related to Zener tunneling and 
Franz-Keldysh effect.

\section{Model and general consideration}

We choose to discuss the electric field effects on tunneling through a spatially 
finite and
energetically restricted band of electronic states in the context of electron
ballistic transport in superlattices. The profile of ${\cal N}$-well 
supperlattice model 
potential is shown by dashed line in Fig.  1.  Due to electron confinement in a 
well, the
well spectrum is quantized.  Due to tunneling between neighboring wells, the
quasi-discrete (in its lower part) spectrum of an isolated well splits into
minibands of allowed electron energies, separated by the energy gaps, which do 
not 
contain electron states.  Since the hight and width of barriers can be
engineered freely within a wide range, it is well possible to have the lowest
miniband separated from the others by an energy interval much exceeding its
band width $E_{\rm bw}$.  An example of this kind is represented in Fig.  1.

Under a restriction that the electrostatic potential drop across the 
superlattice
does not exceed much the magnitude of $E_{\rm bw}$, the description of electron
transport along the superlattice can be started from the Hamiltonian matrix for 
a single band
\begin{equation}
H_{nn'} ={\cal E}\left[n-({{\cal N}+1})/{2}\right]\delta_{n,n'} 
+\delta_{|n-n'|,1},
\end{equation}
where indices $n,n'=\overline{1,{\cal N}}$ number the wells in the superlattice 
(see Fig. 1); ${\cal E} = 
eFa/\beta$; $e$, $F$, and $a$  are, 
respectively, the 
absolute value of electron charge, electric field strength, and periodicity of 
the
superlattice $a=w_{\rm w} +w_{\rm b}$; the energy of electron resonance transfer 
between neighboring wells 
$\beta$ serves as an energy unit; the zero-field energy of the lowest level in 
the non-interacting wells (the electron on-site energy) is set equal to zero. 

The eigenvalues of the Hamiltonian matrix (1) are lying within the
(dimensionless) energy interval (along $E$ axis in Fig.  1) which is equal to 
$E_{\rm bw}+eV$, $E_{\rm
bw}=4$, $eV={\cal E}({\cal N}-1)$.  Excluding classically inaccessible regions
in the $n$-$E$ 'coordinates', one obtains a tilted band of electron
states.  Such a band is shown in Fig.  1 for two voltages $eV<E_{\rm bw}$ and
$eV>E_{\rm bw}$.  Sloped lines on the band diagrams indicate the $n$-$E$ 
(unshaded) 
regions, where the probability to find an electron with the given energy in a 
well 
$n$ (= 1, 2, ..., ${\cal N}$) is
proportional to the tailing part of the corresponding squared eigenfunction of
$H_{nn'}$.  As is seen from the figure, if the applied potential is smaller, 
than 
the miniband width in zero field, $eV<E_{\rm bw}$, the band spectrum can be 
divided 
into es-band and two sls-bands; at higher voltages $eV>E_{\rm bw}$, the es-band 
is 
replaced by WS-band.

Let us assume that from its opposite boundary layers, the superlattice is 
contacted
homogeneously to ideal semi-infinite drain and source electrodes and, that the
interaction between its first (${\cal N}$th) well and the drain (source) lead
can be described by a single parameter ${\cal V}$.  According to the model, the
electrostatic potential energy inside the leads is constant (though different)
and drops from the source to drain by value of $eV$, entirely in the region
occupied by the superlattice, see energy diagrams in Fig.  2.  Suppose also that
the spatial variables of the Hamiltonian of the system
drain-superlattice-source are separable.  Then, the electron energy $E$ can be
broken into a sum of two non-mixing components which correspond to the energy of
electron motion parallel ($E_{\parallel}$) and perpendicular ($E_{\perp}$) to
the electric field, the latter directed along the superlattice.  In such a 
system, an
electron wave, having the energy $E$ and propagating freely, say, in the source
towards the drain, will be transmitted through the superlattice with the 
probability $T(E,{\cal E})$.
Due to the Landauer-B\"uttiker theory,\cite{27,28,29} the transmission
probability is directly related to the current-voltage relation, see, e.g., the
Datta overview.\cite{30}

Probably, for the first time, the model of ballistic electron transport outlined 
above was suggested and solved on the basis of the Green function technique by 
Caroli {\it et
al.}\cite{31} Originally, it was applied to the description of tunnel current in
metal-insulator-metal heterostructures.  To calculate the current, the
Green function has to be found.  In the cited work, however, the latter was not
specified.  Later on, the variations of Caroli {\it et al.}  treatment have
reemerged in a number of physical contexts.  In particular, it has been used to
examine the quantum conductance of molecular wires.\cite{32,33,34}

For the model at hand, the transmission probability $T(E,{\cal E})$ is 
conveniently
expressible in terms of certain matrix elements of the Green functions referred
to the {\it non-interacting} source and drain leads, and the scattering region,
which here is the superlattice.  The miniband width is normally much smaller,
than that of typical contacting leads.  Therefore, in the actual energy
interval, the variation of the lead-related Green functions can be regarded as
negligible in most cases of interest.  Thus, the only energy dependent
quantities remaining in the transmission probability are those referred to the
superlattice, i.e., to the \mbox{Hamiltonian (1)}.  

Combined together, the model assumptions yield\cite{34}
\begin{equation} 
T(E,{\cal E})= \frac{4A^2G^2_{1,{\cal N}}(E)}
{[1+A^2G_\Delta (E)]^2+A^2[G_{1,1}(E)-G_{{\cal N},{\cal N}}(E)]^2
+4A^2G^2_{1,{\cal N}}(E)}, 
\end{equation}
where
$G_\Delta(E)=G_{1,1}(E)G_{{\cal N},{\cal N}}(E)-G^2_{1,{\cal N}}(E)$; and the 
(superlattice) Green function obeys the equation
\begin{equation}
\sum_{n''=1}^{\cal N}\left(E\delta_{n,n''}- H_{nn''}\right)G_{n''n'}(E) = 
\delta_{n,n'}.
\end{equation}
In Eq. (2), the quantity $A\sim{\cal V}^2$ (independent of energy) associates 
with 
the
weighted local density of states on the drain/source-superlattice interfacial
surfaces.  The Green function matrix elements, which appear in the above
equations, depend on $E_{\parallel}$ and ${\cal E}$.  These are the only energy 
dependent
quantities in Eq.  (2) and therefore, here and henceforth the subscript in 
$E_{\parallel}$, as well 
as an indication of the Green function dependence on ${\cal E}$, are
omitted for brevity.  Analytical expressions of 
$G_{11}(E)=-G_{{\cal N}{\cal N}}(-E)$, and $G_{1{\cal N}}(E)$ are given 
by\cite{23}
\begin{equation}
D_{\cal N}(E,{\cal E})G_{11}(E)=
J_{\mu +({\cal N} -1)/2}(z)
Y_{\mu -({\cal N} +1)/2}(z)
-Y_{\mu +({\cal N} -1)/2}(z)
J_{\mu -({\cal N} +1)/2}(z),
\end{equation}
\begin{equation}
D_{\cal N}(E,{\cal E})=
J_{\mu +({\cal N} +1)/2}(z)
Y_{\mu -({\cal N} +1)/2}(z)
-Y_{\mu +({\cal N} +1)/2}(z)
J_{\mu -({\cal N} +1)/2}(z),
\end{equation}
and $D_{\cal N}(E,{\cal E})G_{1{\cal N}}(E)={\cal E}/\pi$, where
$\mu \equiv E/{\cal E}$, $z \equiv 2/{\cal E}$, and $J_{\mu}(z)$ and 
$Y_{\mu}(z)$ are the Bessel functions of the first and second kind, 
respectively. 

Before going into a detailed discussion of $T(E,{\cal E})$ dependence on energy, 
field
strength, and superlattice length (the latter is not indicated explicitly), some 
relevant remarks might be useful.  As is
mentioned above, the $n$-$E$ (length-energy) areas, which are classically 
accessible
for electrons (shaded in Figs.  1 and 2) are distinct, depending on whether the
electrostatic energy is smaller or larger, than the zero-field band width.  For
this reason, the voltages restricted by the conditions $eV<E_{\rm bw}$ and
$eV>E_{\rm bw}$ will be referred as {\it low} and {\it high} voltages,
respectively.  The qualitative difference between the two cases stems from that 
the electron states extended over the entire superlattice (indicated
in the figures by 'es') can exist at low voltages, but not at high voltages, 
when the es-band is shrunken and the WS-band opens instead.

Contrary to extended states, which "connect" the source and drain electrodes,
the electron states located within either of two triangular areas, are separated
from the other side of the superlattice by a triangular barrier, see Fig. 2.  
(The field induced surface
localized states exist if ${\cal E}\ge 12.2/{\cal N}^3$.\cite{25}) Two 
triangular barriers separate the states located within the
WS-band.  Hence, depending on the applied voltage and energy of electrons to be
transmitted through the superlattice, the electron flux can encounter no
barriers at all, one triangular barrier, two triangular barriers, or a trapezoid
barrier, as is exemplified by arrows labeled in Fig.  2 by RT, FN, TB, and WS.
The through trapezoid barrier (TB) tunneling, which also includes the 
Fowler-Nordheim (FN) regime of tunneling, has been
extensively studied in the WKB approximation.\cite{35} Such an approach is
inapplicable for the description of the present discrete, finite band model.  In
particular, it fails to reproduce the resonance structure of $T(E,{\cal E})$,
which might be expected for the sls-assisted tunneling (i.e., tunneling through
a triangular well).  For the resonance tunneling (RT) transmission, the energy
and length dependence of $T(E,{\cal E})$ in zero field ${\cal E}=0$ has been
analyzed previously.\cite{32} However, the
presence of electric field gives rise to new regularities which have not been 
studied so far.  In what follows,
the transmission spectrum will be examined in the RT, FN, WS, and TB energy
intervals, where surprisingly accurate explicit expressions of $T(E,{\cal E})$
can be derived from the model exact equations (2), (4), and (5).

\section{Extended states assisted tunneling}

At low voltages the mid band levels can form WSLs with noncanonical level 
spacing\cite{25}
\begin{equation}
E_l=
\left\{
\begin{array}{c}
l\\
l+1/2
\end{array}\right\}
\frac{1}{1-2m'/m}{\cal E}_{m'/m},
\end{equation}
where the upper (lower) factor should be used for an odd ${\cal N}=2N+1$ (even 
${\cal N}=2N$) number of wells in the superlattice; $l=0,$ 1, 2, ...; $m$ and 
$m'< m/2$ are positive integers, which specify the value of the field parameter
\begin{equation}
({\cal N}-1){\cal E}_{m'/m} = 4 \cos \left ( \pi \frac{m'}{m}\right ).
\end{equation}
For some voltages, the correspondence between the approximate WSL energies and 
exact eigenvalues (solutions to ${\cal D}_{\cal N}(E,{\cal E})=0$) can be seen 
in Fig. 3, where the former and the latter are indicated by circles and stars, 
respectively. In our example, ${\cal N}=101$; the increase/decrease of ${\cal 
N}$ 
improves/worsens the accuracy of Eq. (6).

As is implied by the derivation of Eqs.  (6) and (7),\cite{25} the WSL with the
level spacing, equal to ${\cal E}$ times any but $>1$ integer or ratio of two 
integers, appears in the spectrum mid under a voltage, determined by Eq.  (7),
within the range $0<eV<E_{\rm bw}$.  This result may be thought as a
reminiscence of approximately equidistant levels with the spacing equal to
$\pi/N$ near the center of zero-field band.  For an experimental observation,
the WSLs of the type $E_l=lq{\cal E}$, where $q$ is equal to either a
positive integer $\ge 2$ or $1+ 2m'/(m-2m')$ with a not too small second 
summand, are perhaps most meaningful, because the level spacing is well 
distinguishable from
that in the canonical WSL.  It also may be of importance from the experimental
point of view that any of (noncanonical) WSLs can be obtained only at a unique
voltage.  For instance, the WSL with triple-${\cal E}$ level spacing should 
appear, when the
electrostatic energy is exactly equal to $E_{\rm bw}/2$, if ${\cal N}$ is odd,
or to $E_{\rm bw}/2-{\cal E}$, if ${\cal N}$ is even.  As we
believe, these results can help to identify the electric field effects.

To obtain an explicit expression of $T(E,{\cal E})$, we use in Eq.  (2) an
approximation of the Green function matrix elements at the energies $E_n=n{\cal 
E}$ with $n$ equal to a positive integer restricted by the condition $n<< {\cal 
N}$.  Performing some algebra, which is 
much in the spirit of the derivation of noncanonical WSLs,\cite{25} we get 
instead of Eqs.  (4) and (5)
$$
G_{11}(E_n,{\cal E})
\approx
\frac{\sin[(2n-1)\chi]}
{\sin(2n\chi)},
\eqno{(8{\rm a})}
$$
$$
G_{{\cal N}{\cal N}}(E_n,{\cal E})
\approx
\displaystyle
-\frac{\sin[(2n+1)\chi]}
{\sin(2n\chi)},
\eqno{(8{\rm b})}
$$
$$
G_{1{\cal N}}(E_n,{\cal E})
\approx
\displaystyle
(-1)^{N+n+1}
\frac{\sin \chi}
{\sin(2n\chi)},
\eqno{(8{\rm c})}
$$
where $\chi=\arccos({{\cal E}}{N}/2)$.  Unlike the exact expressions for the 
Green
function matrix elements, Eqs.  (8a), (8b), and (8c) make sense only for large
values of ${\cal N}$ (set to be odd), and only for the indicated
energies.  Using the above expressions in Eq.  (2) we obtain
$$
T(E=n{\cal E},{\cal E})
\approx
\frac{4A^2\sin^2\chi}
{4A^2\sin^2\chi+\sin^2(2n\chi)\left[1+2\cos(2\chi) A^2+A^4\right]},
\eqno{(9{\rm a})}
$$
or, equivalently,
$$
T(n{\cal E},{\cal E})
\approx
\frac{A^2(4-{\cal E}^2N^2)}
{A^2(4-{\cal E}^2N^2)+\sin^2[2n \arccos({\cal E}N/2)]
[(1-A^2)^2+A^2{\cal E}^2N^2]}.
\eqno{(9{\rm b})}
$$
Note that Eq. (9a) is also valid for high voltages, in which case 
$\chi=\cosh^{-1}({{\cal E}}{N}/2)$, and $\sin\chi$ and $\cos\chi$ should be 
replaced by $\sinh\chi$ and $\cosh\chi$, respectively.

According to Eq. (9b), if the applied potential satisfies Eq. (7), 
the transmission coefficient is equal to unit for $E=n{\cal E}$ with $n$ 
divisible by $m$, if $m$ is odd, or by $m/2$, if it is even. This result makes 
it 
obvious that $T(E,{\cal E})$ has a  
resonance-like structure, as a function of the applied voltage at a fixed energy 
and, as a function of energy at a fixed voltage. Thereby, noncanonical WSLs are 
exposed in the transmission spectrum, see Fig. 3, as almost equidistant peaks. 
It is seen that the level spacing is in a good agreement with formula (6) (the 
WSL energies are indicated in Fig. 3 by circles on the E-axis). However, the 
above approximation does not reproduce the decrease of transmission peak maxima, 
which is a pronounced tendency of the exact dependence of the transmission 
probability on energy. Represented by a few graphs in Fig. 3, it reflects a 
general trend of resonance tunneling phenomena, which implies the unit 
transmission probability only for totally symmetric systems (e.g., as 
conventional barrier-well-barrier symmetric heterostructures). In the given 
case, the system symmetry is broken by the applied potential, the increase of 
which results in an increasing suppression of the transmission peaks close to 
the es-band edges.

With the increase of superlattice-to-lead coupling parameter $|A|$, transmission
peaks broaden, see Fig.  3, which is a kind of expected behavior from other
resonance tunneling structures.  Figure 3 also exposes intuitively not expected
behavior of the transmission spectrum response to increasing voltage.  As is
clearly seen from the comparison of the peak positions for ever lager potential
(from bottom to top), the peak spacing {\it decreases} with the {\it increase}
of ${\cal E}$.  So, in addition to a distinctive level spacing, and hence, peak
spacing, which is characteristic for noncanonical WSLs, they are also
distinguishable, from the experimental point of view, due to a specific field
dependence of the transmission peak spacing.  The latter decreases with $eV$,
while in a canonical WSL it should increase.

\setcounter{equation}{9}

\section{Surface localized states assisted tunneling}
By using the standard approximations of Bessel functions
with large arguments and small or large orders,\cite{36} it can be shown
that
the exact expressions of $G_{11}(E)$, $G_{{\cal N}{\cal N}}(E)$, and
$G_{1{\cal
N}}(E)$ [see Eqs. (4) and (5)] are accurately reproduced within the
sls-band
energy interval $|2-eV/2|+{\cal E}<E<2+eV/2-{\cal E}$ by the following
relations
\begin{equation}
\left \{
\begin{array}{c}
{\cal D}_{\cal N}(E,{\cal E})\\ 
{\cal D}_{\cal N}(E,{\cal E})G_{11}(E,{\cal E})\\
{\cal D}_{\cal N}(E,{\cal E})G_{{\cal N}{\cal N}}(E,{\cal E})
\end{array}
\right \}\approx
\frac{{\cal E}/{\pi}}{\sqrt{\sin\xi\sinh\delta}}\exp\left(\frac{2}{\cal
E}\Phi_\delta\right)
\left \{
\begin{array}{c}
\cos\left(
\displaystyle\frac{2}{\cal E}\Phi_{\xi}-\frac{\pi}{4}+\xi\right)
e^{\delta}\\
\cos\left(\displaystyle\frac{2}{\cal
E}\Phi_{\xi}-\frac{\pi}{4}+\xi\right)\\
\cos\left(\displaystyle\frac{2}{\cal
E}\Phi_{\xi}-\frac{\pi}{4}\right)e^{\delta}
\end{array}
\right \},
\end{equation}
where
$
2\cosh\delta = E+ {eV}/{2}$,
$\Phi_\delta=\delta \cosh \delta - \sinh \delta$,
$2\cos \xi = E- {eV}/{2}$ ($0\le\xi\le\pi$), and
$\Phi_\xi= \sin\xi -\xi\cos\xi$. 
Note that the excluded ${\cal E}$ intervals can contain not more, than one sls 
level each.

Using the above expressions for the Green function matrix elements in
Eq. (2),
we get
\begin{equation}
T(E,{\cal E}) =
\frac{4 A^2 \sinh \delta
\sin \xi \exp(-2\delta)}
{(1
 + A^2e^{-2\delta})\left[\cos^2\left(\displaystyle \frac{2}{\cal E}
\Phi_\xi-
\displaystyle\frac{\pi}{4}+\xi\right)+
A^2\cos^2 \left(\displaystyle \frac{2}{\cal E} \Phi_\xi-
\displaystyle\frac{\pi}{4}\right)\right]}
\exp \left(-\displaystyle\frac{4}{{\cal E}}\Phi_\delta \right),
\end{equation}
showing a resonance-like structure modulated by a function $\exp \left 
(-4\Phi_\delta/{\cal E} \right )$ that decays exponentially with the increase of 
energy.

For the energies of sls-band levels $E^{\rm sls}_i$, given by solutions to 
equation ${\cal D}_{\cal N}(E,{\cal E})\sim \cos\left({2}\Phi_\xi/{\cal E}-
{\pi}/{4}+\xi\right)=0$  within the energy interval 
$\overline{|2-eV/2|,2+eV/2}$,\cite{25} expression (11) simplifies to
\begin{equation}
T(E^{\rm sls}_n,{\cal E})=
\frac{4\sinh\delta_n\exp(-2\delta_n)}
{\sin\xi_n\left[1 + A^2\exp(-2\delta_n)\right]}
\exp\left(-\frac{4}{\cal E}\Phi_{\delta_n}\right),
\end{equation}
where $\delta_n = \cosh^{-1}\left(E^{\rm sls}_n/2+ {eV}/{4}\right)$ and
$\xi_n = \arccos\left(E^{\rm sls}_n/2- {eV}/{4}\right)$. In the case of weak 
superlattice-to-lead coupling $A^2<<1$, Eq. (12) 
determines local maxima of $T(E,{\cal E})$. Thus with the replacement $E^{\rm 
sls}_n\rightarrow E$, the r.h.s. of Eq. (12) gives a function enveloping peaks 
of 
the transmission spectrum in the region of sls-assisted tunneling.

As is seen in Fig. 4, except a small ${\cal E}$ interval above the top of es- or 
WS-band, an explicit analytical description of the sls-assisted tunneling given 
by 
Eq. (11) is indistinguishable from exact calculations. This is a central result 
of 
the section. It might be worth emphasizing that details of the resonance 
structure 
of $T(E,{\cal E})$, such as the position of peaks, and their width and 
intensity, 
are model dependent. In contrast, factor $\exp\left(-4\Phi_{\delta}/{\cal 
E}\right)$, which prescribes for the sls-assisted tunneling probability an 
exponential {\it decrease} with the increase of energy (counted from the 
spectrum 
center) and an exponential {\it increase} with the increase of electric field 
strength, is characteristic for the linear drop of the electrostatic potential 
within the scattering region and does not depend on a particular model of the 
interface (connection) with the leads. The reason for this is that sls-assisted 
tunneling is controlled by exponential tailing of the sls wave function in the 
classically forbidden region. Therefore, the exponential decay of the through 
sls-band tunneling probability can be evaluated from the ratio of the 
probabilities 
to find an electron 
with, say, the energy $E^{\rm sls}_n$ in the first and last wells of the 
isolated 
supperlattice. Such an approach, which is much more simple technically (since 
the 
leads and connection to them are out of consideration), gives a reasonable 
result 
even for the pre-exponential factor of the enveloping function\cite{37} 
$T(E^{\rm 
sls}_n,{\cal E})$ =
$\left(\sinh\delta_n/\sin\xi_n\right)
\exp(-2\delta_n)\exp\left(-4\Phi_{\delta_n}/{\cal E}\right)$, which differs from 
the correct expression (12) only by factor of four.

To make Eq. (11) easy readable, one can use an approximate expression 
\begin{equation}
3\Phi_{\delta}\approx
\left\{
\begin{array}{ll}
(E-\overline{eV}/2)^{3/2},& eV\le E_{\rm bw},\\
(E+\overline{eV}/2)^{3/2},& eV>E_{\rm bw},
\end{array}
\right.
\end{equation}
where $\overline{eV}=|eV-E_{\rm bw}|=|eV-4|$ is the excess of electrostatic 
potential energy over zero-field band width, and $\overline{eV}/2$ has the 
meaning 
of the top of es-band (WS-band) for low (high) voltages. The above approximation 
reproduces the exact dependence reasonably well up to $\overline{eV}$ of the 
order 
of $E_{\rm bw}$, see Fig. 5.

Using Eq. (13) in Eq. (11), the exponential factor
$\exp \left(-4\Phi_{\delta}/{\cal E}\right)$ can be replaced, if $eV< E_{\rm 
bw}$, 
by
$\exp \left[-4(E-\overline{eV}/2)^{3/2}/(3{\cal E})\right]$, which is identical 
to 
that appears in the Fowler-Nordheim theory of field emission.\cite{38} At high 
voltages, however, factor 
$\exp \left[-4(E+\overline{eV})^{3/2}/(3{\cal E})\right]$ does not have a 
semi-classical analogy. In some more details, links of Eq. (11) with the 
probability of tunneling through a triangular barrier are discussed in Ref. 37.

\subsection{Effects of coupling on sls transmission spectrum}
The superlattice-to-lead connection (represented in Eq. (11) by a single 
parameter 
$A$) is dependent on a number of factors. For this reason, in relevant 
heterostructures the magnitude of effective coupling $|A|$ may vary by orders. 
It is of interest therefore, to trace the dependence of the sls-band 
transmission spectrum on the coupling strength.

A common expectation is that with the increase of coupling strength the 
resonance 
structure is smeared out. This is really true for resonance tunneling in zero 
field. By contrast, as can be readily seen from Eq. (11), such an expectation is 
not justified for sls-assisted tunneling. In the latter case, the peak position 
and 
their sharpness is essentially determined by two cosine terms in the denominator 
of 
r.h.s. in Eq. (11). In the absence of anyone of the two, the transmission 
spectrum 
would contain infinitely high resonances either at the sls energies $E_n^{\rm 
sls}$ 
($A=0$, the extreme case of weak coupling) or at zeros of 
$\cos\left(2\Phi_\xi/{\cal E}-\pi/4\right)$ (the extreme case of strong 
coupling). 

If the coupling is weak, the peak spacing repeats the level spacing within 
sls-band. In due course, the latter is proportional to ${\cal E}$, 2${\cal E}$, 
and 
ruled by poles of the Airy function, within the 
lower, mid, and upper part of the sls-band, respectively.\cite{25} 
These regularities can thus be observed in the sls-band transmission 
spectrum, as is illustrated in Fig. 4a by open circles (${\cal E}$ spacing), 
filled 
circles ($2{\cal E}$ spacing), and squares (Airy type spectrum). 

In the case of strong coupling $A^2>>1$, the position of peaks in $T(E,{\cal 
E})$ 
is determined by zeros of the second summand of the denominator in Eq. (11), and 
we 
have instead of Eq. (12) 
\begin{equation}
T(E_p,{\cal E})=
\frac{4\sinh\delta_p}
{\sin\xi_p}
\exp\left(-\frac{4}{\cal E}\Phi_{\delta_p}\right),
\end{equation}
where the values of $E_p$ are given by solutions to equation 
$\cos\left({2}\Phi_\xi/{\cal E}-{\pi}/{4}\right)=0$. These are shifted with 
respect 
to sls energies $E_n^{\rm sls}$, and can be shown to obey the same regularities 
as 
those, observed for the eigenvalues. The hight of peaks does not differ much in 
the 
cases of strong and weak coupling, as can be seen from comparison of Eqs. (12) 
and 
(14). At the same time, if $|A|$ is large, the wells between peaks are 
approximately $A^2$ times deeper. Summarizing, in the case of sls-assisted 
tunneling, strong coupling with the leads makes the resonance structure of the 
transmission spectrum even more pronounced. Otherwise, the weak and strong 
coupling 
result in a similar structure of the transmission spectrum with identical 
regularities of the peak spacing dependence on the electric field, see Fig. 6.

On the other hand, it follows from Eq. (11) that the resonance structure will be 
essentially smeared out in the case of intermediated coupling strength, 
when 
$|A|\approx 1$. This unusual behavior of sls-band transmission 
spectrum 
is illustrated by calculations of $T(E,{\cal E})$ for different couplings, 
$|A|<<1$, $|A|=1$, and $|A|>>1$ in Fig. 4b. One can see that with the increase 
of 
coupling strength, the transmission spectrum, at first, losses its resonance 
structure, and then acquires it again with, roughly, interplaced peaks and 
wells, 
and deepened wells.

The above analysis shows that the resonance tunneling assisted by surface
localized states, which appear as a result of the band tilting by the applied
voltage, is characterized by a kind of unique (at least, met not often)
dependence of the transmission spectrum on the coupling with electron reservoirs
from and to which the tunneling occurs.

Some minimal voltage is required for the first sls to appear and hence,
sls-assisted tunneling to be possible.  Further increase of the applied
potential results in sls-band opening up to its maximal width $E_{\rm bw}$.  The
increase of sls-band width is accompanied by the appearance of peaks in
$T(E,{\cal E})$ with the spacing governed by the following regularities.  For
$eV < E_{\rm bw}/2$, the peak spacing is close to that of Airy spectrum.  For
$E_{\rm bw}/2 <eV <E_{\rm bw}$, the Airy type peak spacing gradually changes to
the double ${\cal E}$ peak spacing characteristic for $H_{nn'}$ eigenvalues in
the mid of the maximal-width sls-band.  Finally, for $eV>E_{\rm bw}$, there
exists the third characteristic energy interval (close to the top of WS-band),
where peaks of the transmission probability are ${\cal E}$-spaced.  The latter
spacing is commonly regarded as the WSL trademark.

\section{Wannier-Stark states assisted tunneling}

In the energy interval of bulk states (i.e., in the mid of the full spectrum),
switching from the low to high voltages results even in a more profound
restructuring of the transmission spectrum.  This can be expected since the
es-band, which directly connects the source and drain leads, is replaced by the
WS-band, where the electronic states are localized between two mutually inverse
triangular barriers.  Tunneling through es-band was already discussed in Sec.
3.  Treatment of WS-states assisted tunneling, which refers to the energy
interval $E<\overline{eV}/2- {\cal E}$, is similar to the preceding analysis.

In the present case, however,
large ${\cal N}$, explicit expressions of the Green function matrix elements 
are different for the eigenvalue energies $E_n$ (up to exponentially small 
corrections given by $E_n\approx n{\cal E}$\cite{25}) and for $E\ne E_n$. For 
the latter (and ${\cal N}=2N+1$), we have
\begin{equation}
\left \{
\begin{array}{c}
{\cal D}_{\cal N}(E,{\cal E})\\
{\cal D}_{\cal N}(E,{\cal E})G_{11}(E,{\cal E})\\
{\cal D}_{\cal N}(E,{\cal E})G_{{\cal N}{\cal N}}(E,{\cal E})
\end{array}
\right \}\approx
\frac{(-1)^N{\cal E}\sin\left(\pi E/{\cal E}\right)}
{\pi\sqrt{\sinh\delta\sinh\delta'}}
\exp\left(\frac{2}{\cal E}\Phi^+\right)
\left \{
\begin{array}{c}
 e^{\delta+\delta'}\\
e^{\delta'}\\
-e^{\delta}
\end{array}
\right \},
\end{equation}
where
$
2\cosh\delta' = {eV}/{2}-E$, $\Phi^+=\Phi_{\delta} + \Phi_{\delta'}$, and
$\Phi_{\delta'}=\delta' \cosh \delta' - \sinh \delta'$. 
The use of Eq. (15) in Eq. (2) yields an explicit expression
\begin{equation}
T(E,{\cal E})\approx
\frac{4A^2\sinh\delta\sinh\delta'}
{\sin^2(\pi E/{\cal E})\left\{\left[\exp(\delta+\delta')-A^2\right]^2
+A^2\left(\exp\delta+\exp\delta'\right)^2\right \}}
\exp\left(-\frac{4}{\cal E}\Phi^+\right),
\end{equation}
which provides an accurate description of the tunneling probability within 
WS-band except the above indicated ${\cal E}$ interval and energies close to 
values of $n{\cal E}$, see Fig. 6.
For the eigenenergies, we have instead of Eq. (16)
\begin{equation}
T(n{\cal E},{\cal E})\approx
\frac{4\sinh\delta_n\exp\left[-2\left(\delta_n-\delta'_n\right)\right]}
{\sinh \delta'_n\left[1+A^2 \exp(-2\delta_n)\right]}
\exp\left(-\frac{4}{\cal E}\Phi_n^{-}\right),
\end{equation}
where $\Phi_n^{-}=\Phi_{\delta_n} -\Phi_{\delta'_n}$, 
$2\cosh\delta_n = {eV}/{2}+n{\cal E}$, and $2\cosh\delta'_n = {eV}/{2}-n{\cal 
E}$. 

In the case of weak coupling, Eq.  (17) has the same meaning for the WS-band, as
Eq.  (12) for the sls-band; and under the replacement $n{\cal E}\rightarrow E$
(hence, $\delta_n,\delta'_n,\Phi_n^{-}\rightarrow
\delta,\delta',\Phi^{-}=\Phi_{\delta} -\Phi_{\delta'}$, respectively) the
function $T(E,{\cal E})$ defined in Eq.  (17) envelopes the transmission
spectrum over its local maxima.  The envelope is reproduced equally well by Eq.
(9a) with the replacement of the trigonometric functions, which appear in
expression (9a), by their hyperbolic counterparts.

By analogy, the expression of the transmission coefficient, that follows from 
Eq. 
(16) at the energies
$E_n=(n+1/2){\cal E}$ (for odd ${\cal N}$), may be called minima envelope. Its 
expression is given by
\begin{equation}
T(E,{\cal E})\approx
\frac{4A^2\sinh\delta\sinh\delta'}
{\left[\exp(\delta+\delta')-A^2\right]^2
+A^2\left(\exp\delta+\exp\delta'\right)^2}
\exp\left(-\frac{4}{\cal E}\Phi^+\right).
\end{equation}

The maxima- and minima-enveloping functions shown in Fig. 6 exhibit a striking 
difference in their dependence on energy. The reason for this can be 
exposed by using approximation (13) and its analogue for $\Phi_{\delta'}$ 
\begin{equation}
3\Phi_{\delta'}\approx
(\overline{eV}/2-E)^{3/2},
\end{equation}
the accuracy of which is illustrated in Fig. 5. The exponents of the enveloping 
functions are thus defined simply as a difference ($4\Phi^-/{\cal E}$) and sum 
($4\Phi^+/{\cal E}$) of two FN-type exponents 
$4(\overline{eV}/2+E)^{3/2}/(3{\cal E})$ and 
$4(\overline{eV}/2-E)^{3/2}/(3{\cal E})$.  Notice, Eqs. (13) and (19) meet the 
requirements 
$\Phi^-=0$, 
$\Phi^+ = 2\Phi_\delta^{\rm min}=\overline{eV}^{3/2}/(3\sqrt{2})$ at the mid of 
WS-band $E=0$, 
and $\Phi^+ = \Phi^- =\Phi_\delta^{\rm max}=\overline{eV}^{3/2}/3$ at the top of 
WS-band $E=\overline{eV}/2$. These estimates makes it easily quantifiable a 
huge quantitative difference between maxima- and minima-envelope in the mid of 
WS-band.

An extremely sharp resonance structure, exhibited by the transmission spectrum
of WS-states assisted tunneling, has the same nature as a well-known phenomenon
of resonance tunneling through a barrier-well-barrier structure.  In the given
case, the barriers are of a triangular shape.  The structure is totally
symmetric (and the transmission coefficient is equal or close to unit for odd
and even ${\cal N}$, respectively) only at the spectrum mid $E=0$.  The
increase/decrease of energy strongly suppresses local maxima of electron
transmission because of increasing system asymmetry.  In contrast, because the
total length of the two barriers is independent of energy, the minima envelope
of $T(E,{\cal E})$ depends on energy not that strongly.

As can be concluded from Eqs.  (17) and (18) and is exemplified in Fig.  6,
unlike sls-assisted tunneling, the sharpness of resonance structure in the case
of through WS-band electron transmission, to a large extent, is insensitive to
the coupling strength.  Hence, provided the symmetry conditions are met,
tunneling through the mid part of a tilted band may serve as a nearly ideal
energy filter.  The anomalous sharpness and exponential decrease of the
equidistant peaks also can be regarded as a distinctive signature of Bloch
oscillations in through tilted band tunneling and related phenomena.

\section{Through trapezoid barrier tunneling}

To wind up the discussion, we briefly consider the case of tunneling indicated
in Fig.  2 by TB arrow.  For energies outside the tilted band, $E>2+eV/2$, and
for large ${\cal N}$, small ${\cal E}$, the transmission probability (2) can be
shown to take the form
\begin{equation}
T(E,{\cal E})= \displaystyle 
\frac{16A^2\sinh\alpha\sinh\delta\exp[-2(\alpha+\delta)]}
{1 + A^2 [\exp(-2\alpha)+\exp(-2\delta)] + A^4\exp[-2(\alpha+\delta)]}
 \exp\left[-\frac{4}{\cal E}\left(\Phi_\delta-\Phi_\alpha\right)\right],
\end{equation}
where the definition of $\alpha$ differs from that of $\delta'$ only by 
interreplacement $E\leftrightarrow eV/2$, $2\cosh\alpha = E - eV/2$.
Under the indicated restrictions, the energy and field dependence, given by the 
above relation (solid lines in Fig. 7), is in excellent agreement with the model 
exact Eq. (2), except energies close to the sls-band, where Eq. (20) predicts 
the probability of tunneling which is somewhat different from exact values of 
$T(E,{\cal E})$ (not seen in the scale used).

Equation (20) and particularly, the exponential factor 
$\exp\left[-4\left(\Phi_\delta-\Phi_\alpha\right)/{\cal E}\right]$ looks very 
much 
differently from the usual WKB expression of probability to tunnel through a 
trapezoid barrier. Nevertheless, the WKB result can be retrieved from Eq. (20) 
by passing to the continuous limit, in the same way, as the FN exponential 
factor can be obtained from Eq. (11).\cite{37} Skipping quite boring 
calculations, we present only the final result for the WKB equivalent of Eq. 
(20) (dashed line in Fig. 7)
\begin{equation}
T_{\rm WKB}(E,eV)= \displaystyle \frac{16A^2\sinh^2\delta\exp(-2\delta)}
{[1 + A^2\exp(-2\delta)]^2 }
 \exp\left[-2\delta{\cal N}\left(1-\frac{eV}{4\delta^2}\right)\right].
\end{equation}

In zero field limit, Eq. (20) transforms into 
\begin{equation}
T^{(0)}(E)= \displaystyle \frac{16A^2\sinh^2\delta^{(0)}\exp(-2\delta^{(0)})}
{[1 + A^2\exp(-2\delta^{(0)})]^2 }
 \exp\left(-2\delta^{(0)}{\cal N}\right),
\end{equation}
where $2\cosh\delta^{(0)}=E$. Hence, without the second term in the exponent in 
Eq. (21), the WKB and zero-field expressions of $T(E,{\cal E})$ coincide in 
their functional dependence on $\delta$ and $\delta^{(0)}$, i.e., on the 
imaginary electron wave vector within the barrier. This proves the identity of 
$T_{\rm WKB}(E,eV)$ with Eq. (20) in zero field limit. Moreover,
as could be expected, the WKB expression fairly well describes tunneling through 
a trapezoid barrier, capped by an (energy bounded) tilted band, up to $eV$ 
values comparable with $E_{\rm bw}$, see Fig. 7, but not the case of high 
voltages.

So, Eqs. (20), (11), and (16) give an accurate explicit description of electron 
tunneling spacified in Fig. 2 by TB, FN, and WS arrows, respectively. The latter 
equation corresponds to a purely quantum case and therefore, it does not have a 
semi-classical analogue, as do the two former equations. For the es-band, the 
resonance structure of the transmission spectrum is shown to contain, at certain 
voltages, evenly spaced peaks of noncanonical WSLs the existence of which has 
been predicted in Ref. 25.

\section{Conclusion}

The transmission probability, describing electron ballistic transport between 
two 
leads connected electronically via a single tilted band, is presented as an 
explicit function of electron energy, electric field parameter, thickness of the 
contact (given by a superlattice, dielectric layer, or relevant) and parameter 
of 
lead effective coupling to the contacting region. The derived expressions bring 
to 
light all characteristic dependencies of the tunnel event, making every point, 
one 
would like to know about this particular model of electric field effects on 
tunneling, easy for the understanding. A number of conclusions is made 
throughout 
the discussion and their manifold physical and experimental implications are 
illustrated in various ways.
No doubt that all of them were to an extent present in numerous related studies 
but 
had never been exposed with the present degree of explicitness and completeness 
covering all typical situations consistent with the model. The potential 
application of obtained results and methodology of their derivation is even more 
promising. Next challenge is to describe the Zener tunneling and Franz-Keldysh 
effect in finite, particularly, ultra-thin structures. Work in this direction is 
currently in progress.

\begin{center}
{\bf ACKNOWLEDGMENTS}
\end{center}
Support from the Swedish Research Council is
greatfully acknowledged.

\newpage

\includegraphics{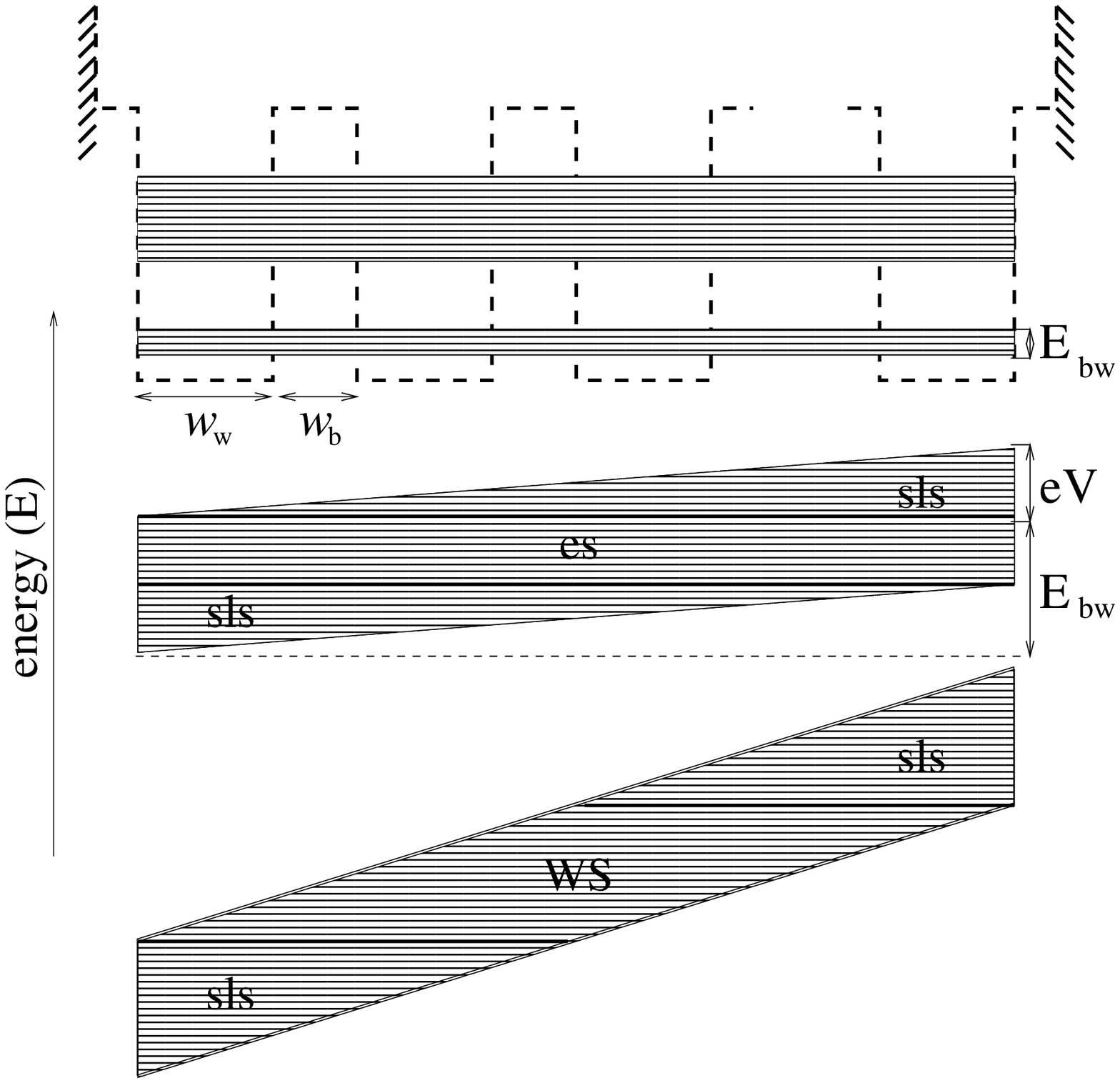}

\vspace*{3.5cm}
{\hspace{1.6cm}\Large{1} \hspace{2.2cm} \Large{2} \hspace{2.2cm}
\Large{3} \hspace{1.2cm} \Large\bf{.\,.\,.} \hspace{.8cm}
\Large${\cal N}$}

\vspace{17cm}
\begin{center}
{\Large Fig. 1}
\end{center}

\newpage
\includegraphics{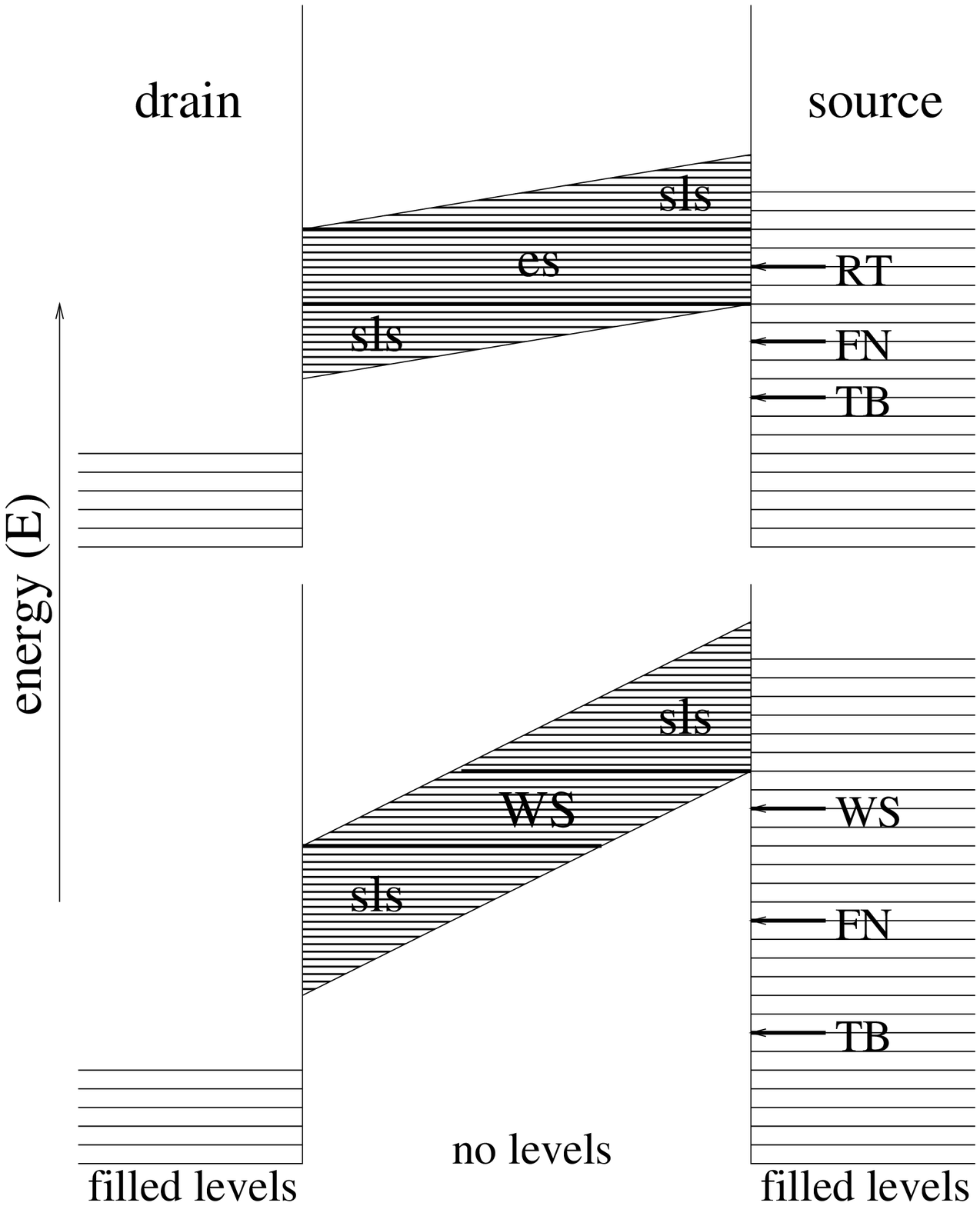}

\vspace*{20cm}
\begin{center}
{\Large Fig. 2}
\end{center}

\newpage

\begin{center}
\includegraphics{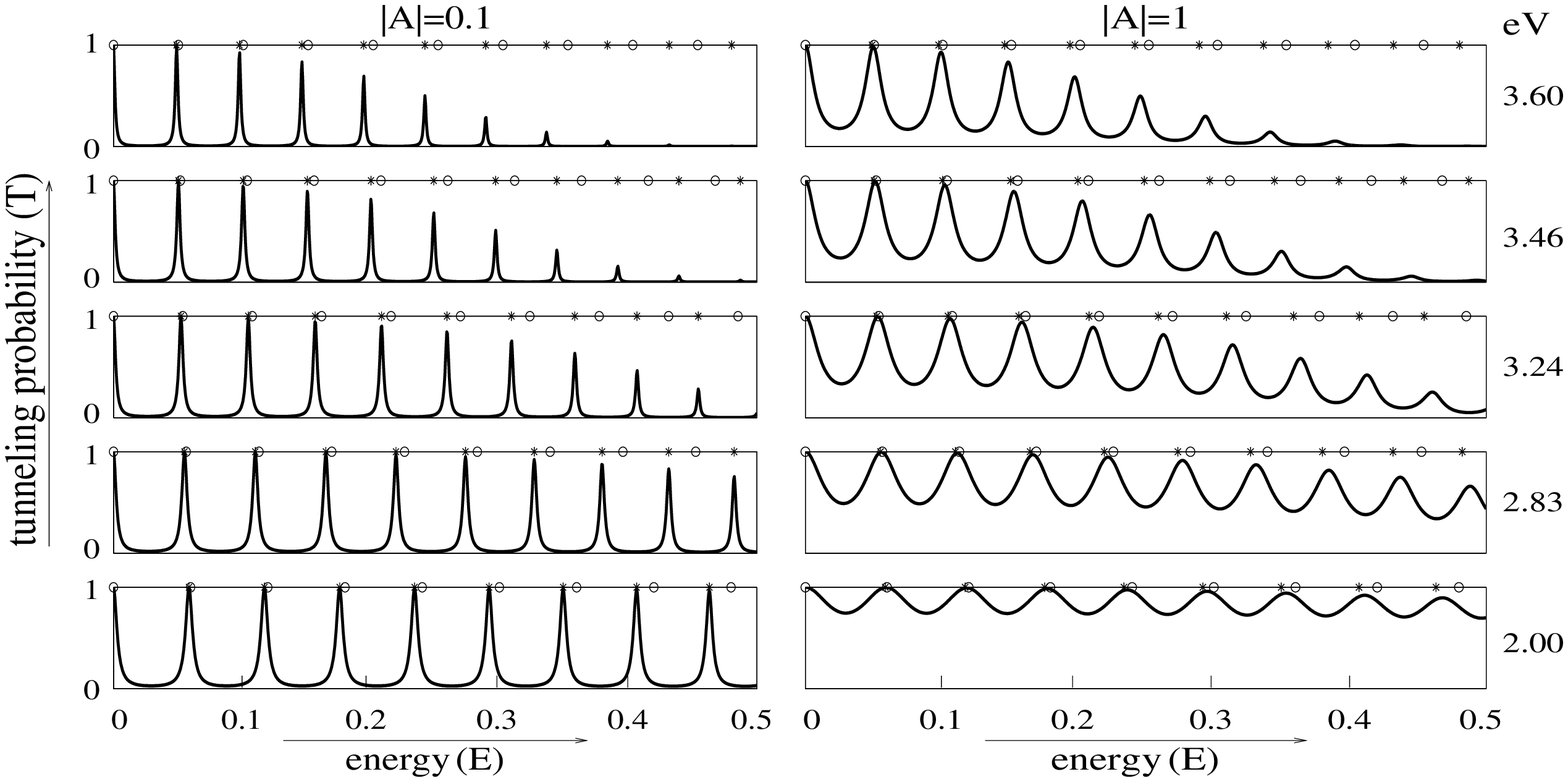}
\end{center}

\vspace{20cm}
\begin{center}
{\Large Fig. 3}
\end{center}

\newpage

\begin{center}
\includegraphics{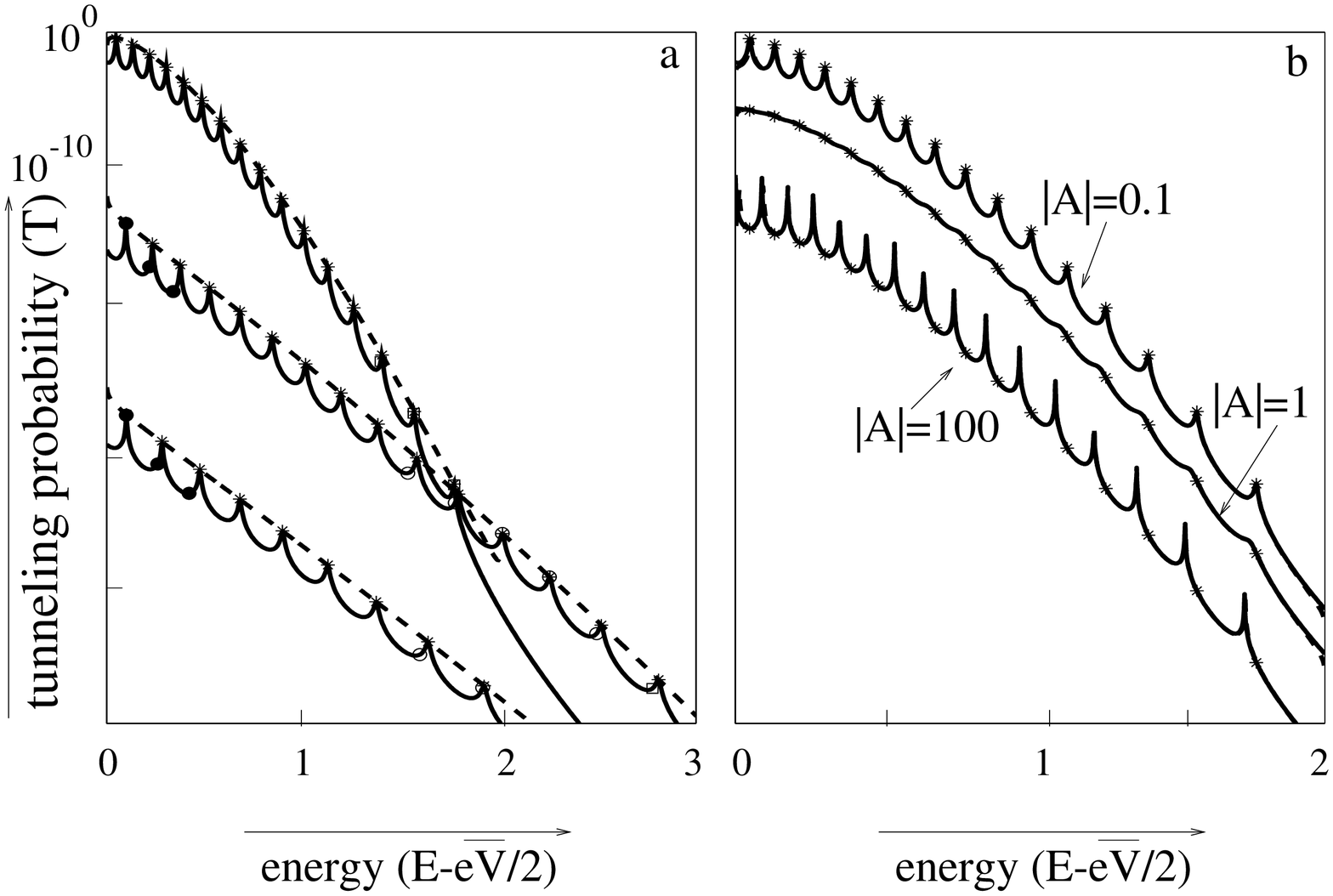}
\end{center}

\vspace{20cm}
\begin{center}
{\Large Fig. 4}
\end{center}

\newpage
\begin{center}
\includegraphics{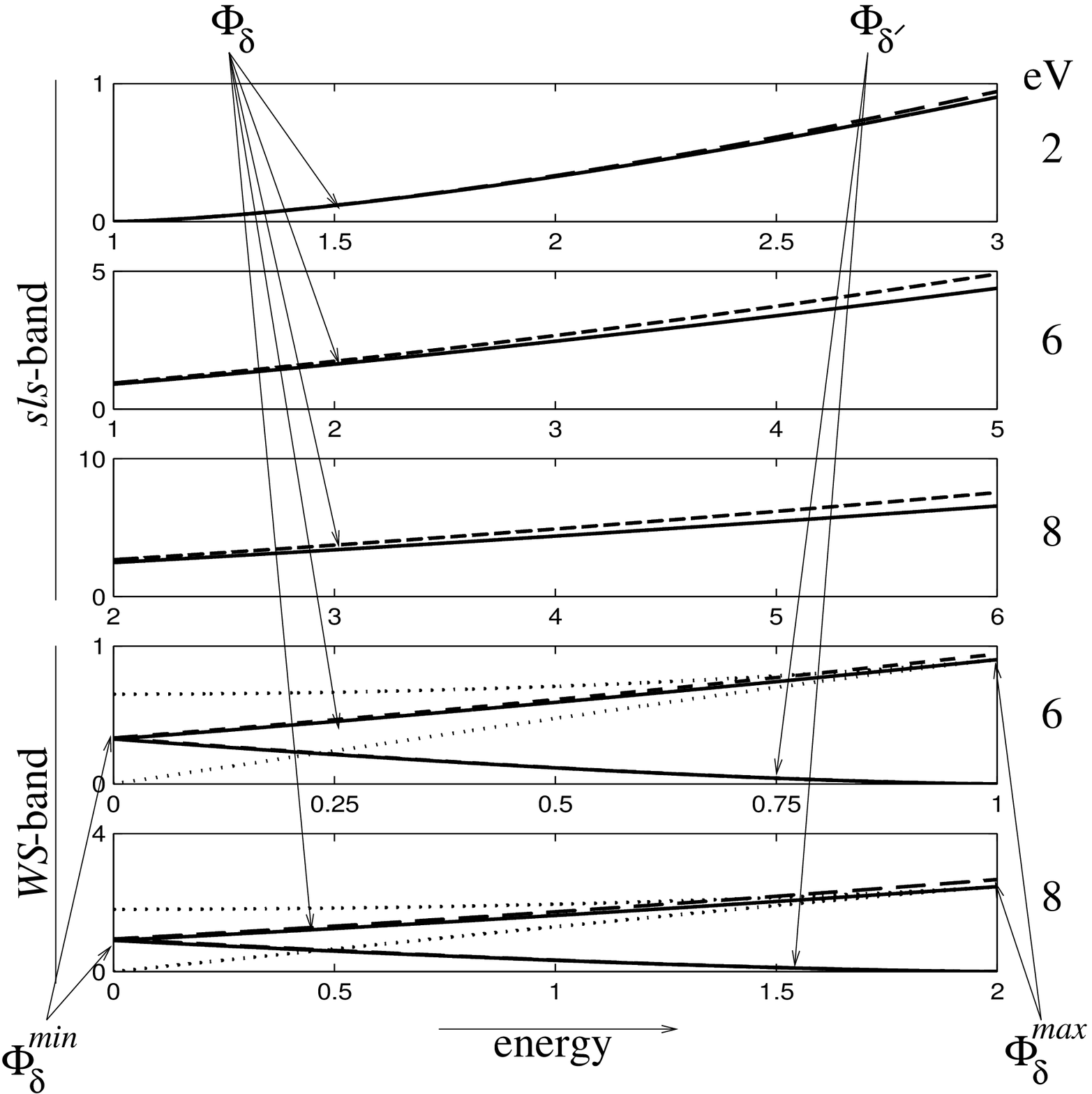}
\end{center}

\vspace{19cm}
\begin{center}
{\Large Fig. 5}
\end{center}

\newpage

\begin{center}
\includegraphics{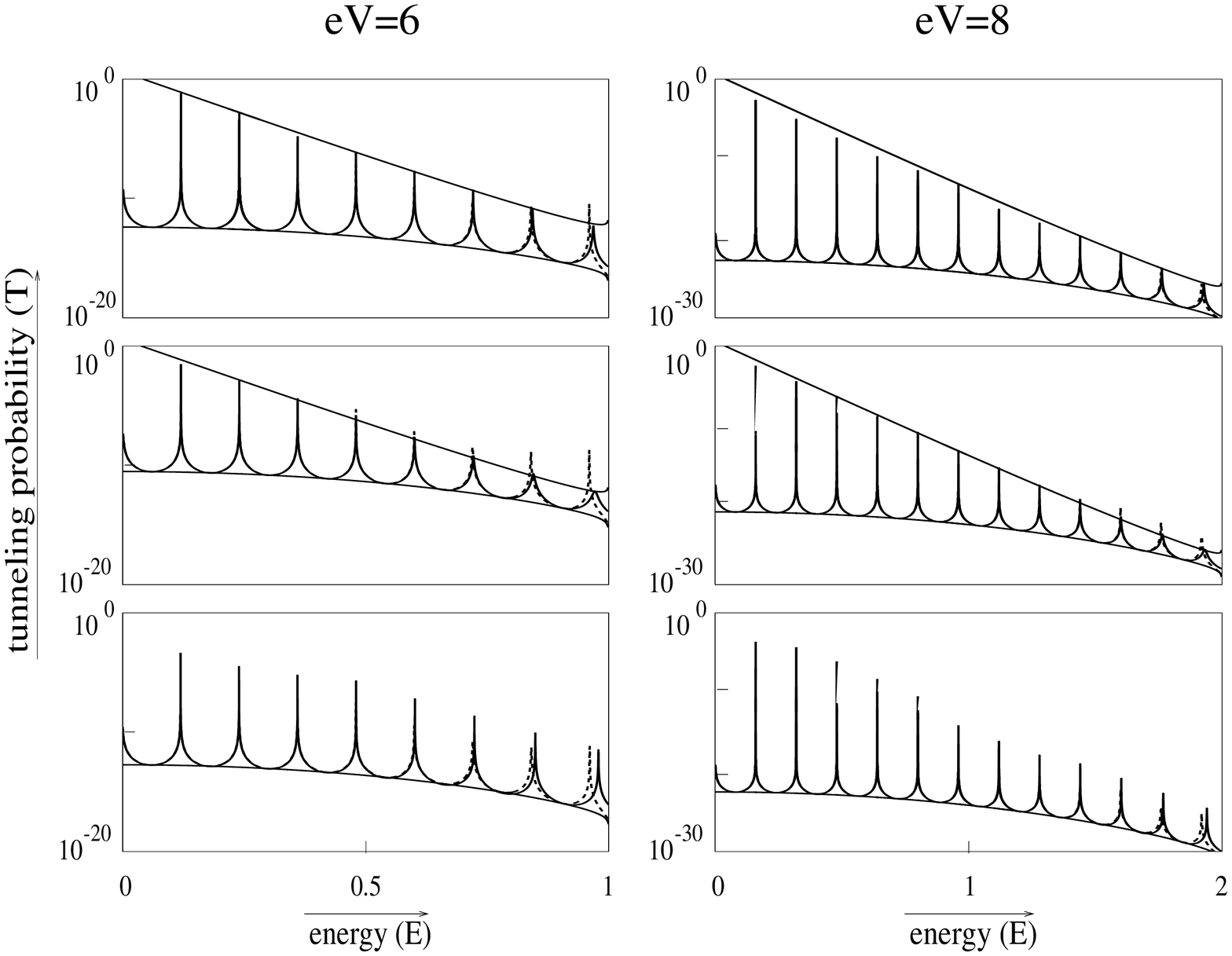}
\end{center}

\vspace{20cm}
\begin{center}
{\Large Fig. 6}
\end{center}

\newpage

\begin{center}
\includegraphics{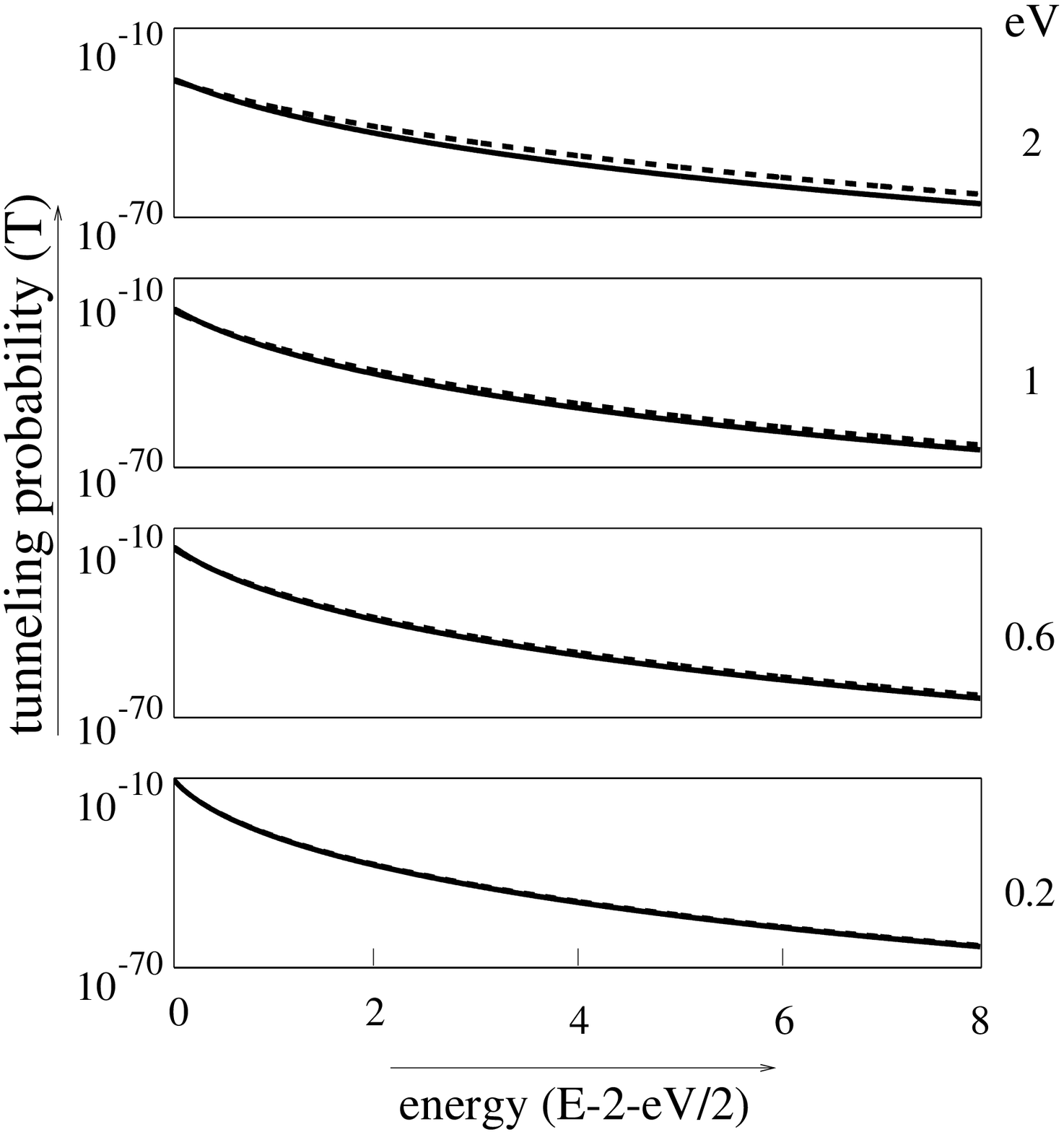}
\end{center}

\vspace{20cm}
\begin{center}
{\Large Fig. 7}
\end{center}

\newpage
\begin{center}
{\Large{\bf Figure caption}}
\end{center}

\noindent
Fig. 1. Upper part: potential energy profile of an ${\cal N}$-well supperlattice 
and its two lowest minibands appearing due to the inter-well tunneling. For a 
superlatice
(Al$_x$Ga$_{1-x}$AsAl$_{x^*}$Ga$_{1-x^*}$As)$_{\cal N}$ with the 
well width $w_{\rm w}$ = 100 \AA, barrier width $w_{\rm b}$ = 40 \AA, barrier 
hight 
= 0.3 eV, and electron 
effective mass = $0.066m_e$, the lowest miniband width (for ${\cal N}>>1$) 
$E_{\rm bw}=3.7$ meV, which  is 24 times smaller, than the band gap.\cite{8}\\
Lower part: The lowest miniband tilted by the applied potential $eV$, 
which is smaller (up) and larger (down) than $E_{\rm bw}$.

\vspace{.5cm}
\noindent
Fig. 2. Energy diagrams of lead-superlattice-lead heterostructure. Electronic 
spectrum of the superlattice is modeled by a single tilted band. At low voltages 
$eV<E_{\rm bw}$ (upper diagram), the electron energy can be tuned to es-band 
(resonance tunneling, RT arrow), sls-band (Fowler-Nordheim type tunneling, FN 
arrow), or it can be out of the tilted band spectrum (through trapezoid barrier 
tunneling, TB arrow). The high voltage case is distinct only by that tunneling 
through the mid part of tilted band is assisted by Wannier-Stark states (WS 
arrow).

\vspace{.5cm}
\noindent
Fig. 3. Miniband transmission spectrum evolution under increasing voltage: from 
bottom to top 
$eV=2.00$, 2.83, 3.24, 3.46, and 3.60; ${\cal N}=101$. The exact level energies, 
i.e., solutions to 
$D_{\cal N}(E,{\cal N})=0$, are indicated by stars. WSLs with noncanonical level 
spacing $3{\cal E}_{1/3}$, $2{\cal E}_{1/4}$, $\frac{5}{3}{\cal E}_{1/5}$, 
$\frac{3}{2}{\cal E}_{1/6}$, and $\frac{7}{5}{\cal E}_{1/7}$ are labeled by open 
circles.

\vspace{.5cm}
\noindent
Fig. 4. Transmission spectrum of sls-assisted tunneling through the lowest 
miniband 
of a 51-well superlattice.\\
a) The miniband is tilted by the electrostatic potential difference $eV$ of (up 
to 
down) $0.5E_{\rm bw}$, $1.5E_{\rm bw}$, and $2E_{\rm bw}$; $|A|=0.1$ (weak 
coupling). Open and filled circles indicate ${\cal E}$- and $2{\cal E}$-spaced 
peaks; and those peaks, which follow the Airy spectrum 
$2+eV/2-E_n =
\left[3\pi\left(n-{1}/{4}\right){\cal E}/2\right]^{2/3}-{\cal E},$\cite{25} are 
indicated by squares. Dashed envelopes represent Eq. (12) with $E^{\rm sls}_n$ 
replaced by $E$.\\ 
b)
$T(E,{\cal E})$ is calculated from Eq. (2) (solid lines) and Eq. (11) (dashed 
line) 
for $eV=0.5E_{\rm bw}$ and $|A|$ = 0.1 (weak coupling), 1 (intermediate 
coupling), 
and 100 (strong coupling). In both figures (a) and (b), stars indicate the 
values 
of $T(E_n^{\rm sls},{\cal E})$.

\vspace{.5cm}
\noindent
Fig. 5. Exact (solid lines) and approximate (dashed lines) dependencies 
$\Phi_\delta(E)$ and $\Phi_{\delta'}(E)$ as they are represented in the text. 
Function $\Phi_\delta(E)$ is specified in Eqs. (10) (exact) and (13) 
(approximate); 
$\Phi_{\delta'}(E)$ is represented in Eq. (15), and its approximate expression 
is 
given by Eq. (19).
Three upper graphs correspond to 
sls-band. In two lower graphs for WS-band, the rising and lowering lines 
represent 
$\Phi_\delta(E)$ and $\Phi_{\delta'}(E)$, respectively;
$3\Phi_\delta^{\rm min}$ = $(\overline{eV}/2)^{3/2}$, $3\Phi_\delta^{\rm 
max}$=$(\overline{eV})^{3/2}$.
Functions $\Phi^\pm$ = $\Phi_\delta\pm\Phi_{\delta'}$ are plotted by dotted 
lines.

\vspace{.5cm}
\noindent
Fig. 6. Transmission spectrum of WS-states assisted tunneling through the lowest 
miniband of a 51-well superlattice. Solid and dashed oscillating lines represent 
exact and approximate expressions of $T(E,{\cal E})$ given in Eqs. (2) and (16), 
respectively. Maxima envelope corresponds to Eq. (17) with $n{\cal 
E}$ replaced by $E$, and minima envelope corresponds to Eq. (18). In upper part 
$|A|=0.1$; in the mid, $|A|=1$; and for two lower graphs, $|A|=100$.

\vspace{.5cm}
\noindent
Fig. 7. Tunneling probability above (or below) the tilted band, calculated from 
Eqs. (20) and (21) for different voltages (as indicated), is represented by  
solid and dashed lines, respectively. In calculations, ${\cal N}=51$, $|A|=0.1$. 


\begin{references}

\bibitem{1} 
G.H. Wannier, Phys. Rev. {\bf 117} (1960) 432.

\bibitem{2} 
G.H. Wannier, Rev. Mod. Phys. {\bf 34} (1962) 645.


\bibitem{3} 
E.E. Mendez and G. Bastard, Physics Today {\bf 46} ({\bf 6}), 34 (1993).

\bibitem{4} 
F. Rossi, Semicond. Sci. Technol. {\bf 13} (1998) 147.

\bibitem{5} 
K. Leo, Semicond. Sci. Technol., {\bf 13} (1998) 249.

\bibitem{6} 
D.H. Dunlap and M.V. Kenkre, Phys. Rev. B {\bf 25} (1986) 3625. 

\bibitem{7} 
M. Holthouse, Phys. Rev. Lett. {\bf 69} (1992) 351. 

\bibitem{8} 
M. Holthouse, Z. Phys. B {\bf 89} (1992) 251. 

\bibitem{9} 
M. Holthouse and D. Hone, Phys. Rev. B {\bf 69} (1993) 6499. 

\bibitem{10} 
S.R. Wilkinson, C.F. Bharucha, K.W. Madison, Q. Niu, and M.G. Raizen, Phys. Rev. 
Lett. {\bf 76} (1996) 4512. 

\bibitem{11} 
Q. Niu, X.-G. Zhao, G.A. Georgakis, and M.G. Raizen, Phys. Rev. Lett. {\bf 76} 
(1996) 4504. 

\bibitem{12} 
M.G. Raizen, C. Salomon, and Q. Niu, Physics Today {\bf 50} ({\bf 7}), 30 
(1997).

\bibitem{13} 
K.W. Madison, M.C. Fischer, R.B. Diener, Q. Niu, and M.G. Raizen, Phys. Rev. 
Lett. {\bf 81} (1998) 5093. 

\bibitem{14} 
K.W. Madison, M.C. Fischer, and M.G. Raizen, Phys. Rev. A {\bf 60} (1999) R1767. 

\bibitem{15} 
M. Gl\"uck, M. Hankel, A.R. Kolovsky, and H.J. Korsch, Phys. Rev. A {\bf 61} 
(2000) 061402(R). 

\bibitem{16} 
J. Zak, Phys. Rev. Lett. {\bf 20} (1968) 1477. 

\bibitem{17}
K. Hacker and G. Obermair, Z. Physik {\bf 234} (1970) 1.

\bibitem{18}
G.C. Stey and G. Gusman, J. Phys. C: Solid State Phys. {\bf 6} (1973) 650.

\bibitem{19}
M. Saitoh, J. Phys. C: Solid State Phys. {\bf 6} (1973) 3255.

\bibitem{20}
H. Fukuyama, R.A. Bari, and H.C. Fogedby, Phys. Rev. B {\bf 8} (1973) 5579. 

\bibitem{21}
Yu.B. Gaididei and A.A. Vakhnenko, Phys. stat. sol. (b) {\bf 122} (1984) 239.

\bibitem{22} 
A.I. Onipko, L.I. Malysheva, and Yu.A. Klimenko, Physica B 
{\bf 225} (1996) 125.

\bibitem{23}
L.I. Malysheva, Ukr. Fiz. Zh. {\bf 45} (2000) 1475.

\bibitem{24}
V.M. Yakovenko and H.-S. Goan, Phys. Rev. B {\bf 58} (1998) 8002.

\bibitem{25}
A. Onipko and L. Malysheva, Solid State Commun. {\bf 118}, 63 (2001); Phys. Rev. 
B {\bf 63} (2001).

\bibitem{26} 
D.L. Smith and S.M. Kogan, Phys. Rev. B {\bf 54}, 10354 (1996), and references 
therein. 

\bibitem{27} 
R. Landauer, IBM J. Res. Develop. {\bf 1}, 323 (1957); Philos. Mag. 
{\bf 21} (1970) 683.

\bibitem{28} 
M. B\"uttiker, Y. Imry, R. Landauer, and S. Pinhas, Phys. Rev. B 
{\bf 31} (1985) 6207. 

\bibitem{29} 
A.D. Stone and A. Szafer, IBM J. Res. Develop. {\bf 32} (1988) 384. 

\bibitem{30} 
S. Datta, {\it Electronic Transport in Mesoscopic Systems} 
(Cambridge University Press, Cambridge, 1995).

\bibitem{31} C. Caroli, R. Combescot, P. Nozieres, and D. Saint-James,
J. Phys. C: Solid State Phys. {\bf 4} (1971) 916.

\bibitem{32} 
V. Mujica, M. Kemp, and M. A. Ratner, J.  Chem.  Phys. {\bf 
101}, 6849; {\bf 101} (1994) 6856.

\bibitem{33}
 A. Onipko, Phys. Rev. B {\bf 59} (1999) 9995.

\bibitem{34}
A. Onipko, Yu. Klymenko, and L. Malysheva, Phys. Rev. B {\bf 62} (2000) 10480.

\bibitem{35}
J.G. Simmons, J. Appl. Phys. {\bf 34} (1963) 1793.

\bibitem{36}
M. Abramowitz and I.A. Stegun, {\it Handbook of Mathematical Functions} (New 
York: Dover, 1965).

\bibitem{37}
A. Onipko and L. Malysheva, Proceedings of Molecular Electronics 2000, Kona, 
Hawaii, December 10--14, 2000, to be published.

\bibitem{38}
R.H. Fowler and L. Nordheim, Roy. Soc. Proc. A {\bf 119}(1928) 173.


\end{references}
\end{document}